\documentclass{acmart}

\usepackage{lipsum}
\usepackage{xcolor}
\usepackage{listings}
\usepackage{svg}
\usepackage{amsmath}

\usepackage{xpatch}

\usepackage{siunitx}
\definecolor{keyword}{HTML}{2771a3}
\definecolor{pattern}{HTML}{b53c2f}
\definecolor{string}{HTML}{be681c}
\definecolor{relation}{HTML}{7e4894}
\definecolor{variable}{HTML}{107762}
\definecolor{comment}{HTML}{8d9094}

\usepackage[normalem]{ulem}
\newcommand{\keepcomment}{0 } 
\ifnum\keepcomment=1
	\usepackage[colorinlistoftodos,textsize=scriptsize]{todonotes} 
    \newcommand{\stkout}[1]{\ifmmode\text{\sout{\ensuremath{#1}}}\else\sout{#1}\fi}
    
    \newcommand{\todoi}[1]{\todo[inline]{#1}}
\else
	
	\usepackage[disable]{todonotes} 
	\newcommand{\todoi}[1]{\leavevmode\ignorespaces\unskip}
\fi

\definecolor{color1}{rgb}{0.0,  0.0,0.6}
\definecolor{color2}{rgb}{0.29, 0,  0.29}
\definecolor{color3}{rgb}{0.25, 0.5,0.5}

\makeatletter
\lstdefinelanguage{url}{
    moredelim=[is][\ProcessAmpersand]{\&}{=},
    moredelim=[is][\ProcessQM]{?}{=},
    moredelim=[l][\color{color1}]{http://},
    alsoletter={0,1,2,3,4,5,6,7,8,9,.,/,:},
    showstringspaces=true,
    identifierstyle=\color{color3},
    literate = |{{\textcolor{color1}{|}}}1,
}

\def\ProcessAmpersand%
{%
    \lst@CalcLostSpaceAndOutput%
    \textcolor{color1}{\&}%
    \color{color2}%
    \aftergroup\ProcessClosingEq%
}

\def\ProcessQM%
{%
    \lst@CalcLostSpaceAndOutput%
    \textcolor{color1}{?}%
    \color{color2}%
    \aftergroup\ProcessClosingEq%
}

\def\ProcessClosingEq{\textcolor{color1}{=}}

\begin{document}

\newcommand{\mytitle}{\emph{PoolLines}: Modeling Carpooling as Ephemeral Lines in GTFS for effective integration with Public Transit}

\title[\mytitle]{\mytitle}


\author{Youssef Chaabouni}
\email{youssefchaabouni98@gmail.com}
\affiliation{%
 \institution{Télécom SudParis, IP Paris}
\country{France}
}

\author{Andrea Araldo}
\email{araldo@imtbs-tsp.eu}
\affiliation{%
  \institution{Télécom SudParis, IP Paris}
  \country{France}
}

\author{André de Palma}
\email{andre.de-palma@cyu.fr }
\affiliation{%
  \institution{CY Cergy Paris Université}
  \country{France}
}

\author{Souhila Arib}
\email{souhila.arib@cyu.fr}
\affiliation{%
  \institution{CY Cergy Paris Université}
  \country{France}
}

\renewcommand{\shortauthors}{Araldo et al.}

\lstdefinelanguage{cypher}
{
	morekeywords={
		MATCH, OPTIONAL, WHERE, NOT, AND, OR, XOR, RETURN, DISTINCT, ORDER, BY, ASC, ASCENDING, DESC, DESCENDING, UNWIND, AS, UNION, WITH, ALL, CREATE, CALL, YIELD, DELETE, DETACH, REMOVE, SET, MERGE, SET, SKIP, LIMIT, IN, ON, CASE, WHEN, THEN, ELSE, END, INDEX, DROP, UNIQUE, CONSTRAINT, EXPLAIN, PROFILE, START,
	}
}

\newcommand{\mycdots}{\cdot\!\cdot\!\cdot}
\lstset{language=cypher,
	literate=*
	{...}{$\mycdots$}{1}
	{theta}{$\theta$}{1}
}

\todoi{aa: Add the computation how much CO2 we save.
You take the riders that were not served in the Current System and that are served in the Integrated System. Tu fais la somme des Km q'il feraient s'ils prenaient la voiture privée (en prenant en considération la circuity). Tu soustrais de cette somme le total des Km parcourus par le riders pour faire de détours ``effectifs''. Le résultat te donne le Km gagnés par le système. Écrit ce résultat stp. Multiplie ces Km gagnés par 97 g (en précisant que tu tire la mesure de \cite{ADEME2022}. Comme tes mesure sont sur 30 min, convertissent-les en heure, en multipliant par 2.
}
\newpage
\begin{abstract}
In carpooling systems, a set of drivers owning a private car can accept a small detour to pick-up and drop-off other riders. However, carpooling is widely used for long-distance trips, where rider-driver matching can be done days ahead. Making carpooling a viable option for daily commute is more challenging, as trips are shorter and, proportionally, the detours tolerated by drivers are more tight. As a consequence, finding riders and drivers sharing close-enough origins, destinations and departure time is less likely, which limits potential matching.

In this paper we propose an \emph{Integrated System}, where carpooling matching is synchronized with Public Transit (PT) schedules, so as to serve as a feeder service to PT in the first mile. Driver detours are proposed towards PT selected stations, which are used as consolidation points, thus increasing matching probability. We present a computationally efficient method to represent PT schedules and drivers trajectory in a single General Transit Feed Specification database, which allows to compute multimodal rider journeys using any off the shelf planners. We showcase our approach in the metropolitan area of Portland, Oregon, considering 8k randomly generated trips. We show the benefits of our Integrated System. We find that 10\% more riders find a feasible matching with respect to the status quo, where carpooling and PT are operated separately. We release our code as open source.\footnote{\url{https://github.com/YoussefChaabouni/Carpooling}}

\end{abstract}

\ccsdesc[500]{Applied computing~Transportation}

\keywords{Transportation, Carpooling, Public Transit, GTFS, Open Data }

\maketitle

\section{Introduction}
Mass Public Transit (PT) is irreplaceable to enable mobility in big urban conurbations. However, the level of service offered by PT is generally not satisfactory in the entire urban region~\cite{Badeanlou2022}. Indeed, in suburban areas PT agencies cannot guarantee high frequency and high coverage service, as the lower demand density would determine an unbearable cost-per-passenger. Flexible modes, which adapt their routes on the trip request observed at a certain moment~\cite{Araldo2019a}, are a valid complement to conventional fix PT in such areas. Demand-responsive buses have been mostly considered~\cite{Calabro2021}. On the other hand, the benefits of integrating carpooling to conventional PT have been less studied. Some articles proposing such integration are \cite{stiglic2015benefits,Fahnenschreiber2016,araldo2022pooling}. The first assumes that riders do not necessarily choose the most convenient path but obey to what the system dictates and that carpooling can only occur in the first mile (from the origin to a PT stop) and not in the last mile (from a PT stop to a destination). The second can only match one rider per driver. The third can only handle one PT line only. None of the three has been applied to a real PT network.

In this work, we propose an open-data driven approach to model an \emph{Integrated System}, where conventional fixed PT is complemented by carpooling. Our approach has GTFS data at the core, a standard specification of PT schedules, which allows us to apply it to real PT networks. Our simple key idea consists in representing a carpooling driver trip as an \emph{ephemeral bus line}, which passes only once. We call all such lines \emph{PoolLines}. The advantage of representing drivers as PoolLines into the GTFS database is that the computation of multimodal rider trips, composed of possibly multiple carpooling and PT transit segments, can be done using off the shelf routing software (we use OpenTripPlanner in our case~\cite{OTP,OTPRouting}).

We summarize our contribution as follows:

\begin{itemize}
    \item We propose a modeling approach to describe PT schedules and driver journeys in a single GTFS database~\cite{gtfs:website}.
    \item We compute rider journeys on such a database, using available open-source software. Such journeys can include a chain of carpooling trips with different drivers as well as a combination of carpooling and transit.
    \item We showcase our approach in the metropolitan area of Portland, Oregon.
    \item We show that our Integrated System is able to serve more riders by requiring minimum overall detours to drivers, confirming the importance of using stations as consolidation points.
\end{itemize}


\section{Model and implementation}
\label{sec:model}


\begin{table}[t]
    \centering
    \begin{scriptsize}
    \setlength{\tabcolsep}{2.5pt}
    \begin{tabular}{|l|l|l|l|}
    \hline
        \textbf{Symbol} & \textbf{Description} &
        \textbf{Value}
    \\
    \hline
        $\tau$ & 
        Maximum detour ratio
        & 15\%
    \\
    \hline
        $dw$ & Dwell time (\S\ref{eq:driver-journey})
        & 1 minute
    \\
    \hline
        $d\in\mathcal{D}$ & Drivers &
    \\
    \hline
        $r\in\mathcal{R}$ & Riders &

    \\
    \hline
        $J(d)=\{b_d^0=b_d^\textit{org},b_d^1,b_d^2\dots,b_d^{K_d}=b_d^{\textit{dst}}\}$
        & Journey &
    \\
    \hline
    \end{tabular}
    \caption{Notation and numerical values. }
    \label{tab:notation}
    \end{scriptsize}
\end{table}

Table~\ref{tab:notation} provides the key notations and values used henceforth.

We consider a set of $r\in\mathcal{R}$ riders, each departing at time $t_r$ from origin $\textit{org}_r$ and willing to arrive at destination $\textit{dst}_r$ as soon as possible. Similarly, we consider a set of drivers $d\in\mathcal{D}$, each characterized by $t_d,\textit{org}_d, \textit{dst}_d$.
We denote with $\mathcal{PT}$ public transit schedule, with physical stops $\mathcal{S}$. A scenario is determined by $(\mathcal{R},\mathcal{D}, \mathcal{PT})$.

Given a certain scenario, our Integrated System first computes the journeys of drivers $\mathcal{J}^d=\{J(d)|d\in\mathcal{D}\}$. A journey is a sequence of \emph{stoptimes}
\begin{align}
\label{eq:driver-journey}
J(d)=\{b_d^0=b_d^\textit{org},b_d^1,b_d^2\dots,b_d^{K_d}=b_d^{\textit{dst}}\}.
\end{align}
Each stoptime $b_d^i$ is a pair $(p,t)$, where $p$ is a physical location and $t$ is a time-instant. Each journey must have at least two stop times. The first stoptime $b_d^\textit{org}=(p,t)$ must always be located in the origin $p=\textit{org}_d$ and happen at the departure time $t=t_d$ of the driver. The last stoptime $b_d^\textit{dst}=(p',t')$ must be located at the destination of the driver, $p'=\textit{dst}_d$. If a drive journey has only two stoptimes, it means the driver does not execute any detour. Otherwise, the intermediate stoptimes describe the entire detour.

After computing $\mathcal{J}^d$, the system computes the journeys of riders $\mathcal{J}^r$ exploiting connections via carpooling with drivers or PT. Multiple riders can share a carpooling trip with the same driver. The same rider is also allowed to execute a sequence of carpooling trips with different drivers. \S\ref{sec:journeys} explains how $\mathcal{J}^d$ and $\mathcal{J}^r$ are computed.

\subsection{GTFS model}
\label{sec:GTFS}
We now briefly describe the GTFS specification, on which $\mathcal{PT}$ is modeled~\cite{Fortin2016,gtfs:website}. PT schedules are specified as a set of files. The available bus, subway, train or tramway lines (possibly with different service pattern each) are called \emph{routes}. A route is a sequence of \emph{stops}. The same route is served multiple times, each representing a \emph{trip} departing at a specific time. The event of a trip arriving and departing at a specific stop is called \emph{stoptime}. A trip is thus a sequence of stoptimes. 

The basic idea of this work simply consists in adding driver journeys as new lines into GTFS data, which we call \emph{PoolLines}. For a driver $d$, we add a new route named {\footnotesize{\texttt{	route of carpooler number [id\_of\_carpooler]}}}, a single trip named {\footnotesize{\texttt{	1162238700[id\_of\_carpooler]}}}. We also add two stops:  {\footnotesize{\texttt{DRIVER\_origin\_[id\_of\_carpooler]}}} corresponding to origin $\textit{org}_d$ and {\footnotesize{\texttt{DRIVER\_destination\_[id\_of\_carpooler]}}} corresponding to  destination $\textit{dst}_d$.  For each stop time $b_d^i=(p^i,t^i)$ in the journey (see~\eqref{eq:driver-journey}), we add a corresponding stoptime with location $p^i$ and time $t^i$.The trip is thus a sequence of stoptimes corresponding to~\eqref{eq:driver-journey}. In other words, we represent a driver journey as a bus that passes only once.

\subsection{Computation of driver journeys}
\label{sec:journeys}
We select a subset $\mathcal{M}\subseteq\mathcal{S}$ of physical stops as potential \emph{meeting points}: in such stops, drivers can pick-up or drop-off riders. Such selection is a design parameter. It is advisable to select stops that are well connected to the rest of PT network. Indeed, by dropping-off riders there, they can reach many destinations from there. On the contrary, selecting a stop from which only one PT  line passes with low frequency would be of little use. In our numerical result we select all subway stops as potential meeting points. We will investigate efficient criteria to select meeting points in our future work.

Over the time, new drivers declare their trip to the system, specifying origin, destination and departure time. We assume such declaration occurs some time before the departure time.
For any new driver $d$, we initialize her journey as $d$ to $J(d)=\{b_d^\textit{org},b_d^\textit{dst}\}$. 
Let us denote with $l(J(d))$ the length of journey $J(d)$ in Km and with $l_d^0$ the length of her initial journey $J(d)=\{b_d^\textit{org},b_d^\textit{dst}\}$, without any detour.
The modification of the journey to include detours is similar to our previous work~\cite{araldo2022pooling}. With 50\% probability we first try to add a detour close to $\textit{org}_d$, otherwise we first try close to $\textit{dst}_d$. When trying to add a detour close to $\textit{org}_d$, we take the meeting point $m\in\mathcal{M}$ closest to $\textit{org}_d$ and we compute the time $t$ at which the driver can be there (the details of driver vehicle movement are described in \S\ref{sec:numerical}) including a fix dwell time $dw$, i.e., the time to left riders alight or board. We build a stoptime $b=(m,t)$ and we check whether a new journey $J'(d)=\{b_d^\textit{org},b_d^\textit{dst}\}$ would be acceptable, which depends on whether such a new journey imposes a too long detour. If  $\frac{l(J'(d))-l_d^0}{l_d^0}\le \tau$, where $tau$ is a designed parameter, the new journey is acceptable and we set $J(d):=J'(d)$. Otherwise $J(d)$ remains unchanged. We then try to add a detour close to the destination of the driver, in a similar way, checking that we do not exceed the original length more than $\tau$ times. With the other 50\% probability we would instead first attempt to add a detour close to driver's destination and then close to her origin.

We assume a centralized system controller receiving all drivers' declarations and computing their journeys. For simplicity, we assume drivers always obey to the journeys computed by the controller. Appropriate economic compensation, in order for such an assumption to hold, is outside the scope of this paper and is object of our future work. Once a driver declaration arrive, we assume the controller's computation of her journey is instantaneous. In the time window between the arrival time $at_d$ of driver $d$ declaration and her departure $t_d$, the controller has the opportunity to match riders to $d$. At time $t_d$, just before starting the actual movement, driver $d$ checks what are the stop times that are ``useful'', i.e., the ones in which some rider is picked-up or dropped-off. The driver skips all other stoptimes. 

We add the resulting drivers' journeys $\mathcal{J}^d$ into the GTFS files, as described in \S\ref{sec:GTFS}, in the form of \emph{PoolLines}.

\subsection{Computation of riders' journeys}
\label{sec:rider-journey}
We give the modified GTFS files as input to OpenTripPlanner~\cite{OTP}, an open source software which computes multiple shortest paths from an origin, to a destination at a given departure time. OTP is able to ingest PT schedules in GTFS format and to compute time-dependent paths (the path returned changes depending on the departure time, based on the PT schedule), combining walk and one or more PT lines in a single path. Since some of the GTFS lines are our PoolLines, OpenTripPlanner is able to return paths that include one or multiple carpooling segments.
\todo{aa: Dans la Fig.~1, élimine stp la barre standard du navigateur. Aussi fais plus de zoom (``ctrl'' + ``+''), pour la rendre tout un peu plus lisible.}

When a rider $r$ requests a trip from origin $\textit{org}_r$, destination $\textit{dst}_r$ and departure time $t_r$ (such request may arrive directly at time $t_r$ or in advance), the following HTTP query is sent to the OTP server:

\lstset{language=url}
\begin{lstlisting}
http://localhost:8803/otp/routers/current/plan?
?fromPlace="+str(origin[0])+"%2C"+str(origin[1])
&toPlace="+str(destination[0])+"%2C"+str(destination[1])+"
&time="+str(hours)+"%3A"+str(minutes)+str(am_or_pm)+
&date=07-20-2022
&numItineraries=10
&mode=TRANSIT%2CWALK&maxWalkDistance=2500.032
\end{lstlisting}

\paragraph{}
Here, origin[0] and origin[1] correspond to the longitude and latitude coordinates of the generated rider. We also have the same case for destination[0] and destination[1]. Hours, minutes and am\_or\_pm are variables that express his departure time. We have chosen 10 itineraries so that a single rider can have different options. 
\begin{figure}
    \centering
    \includegraphics[width=\linewidth]{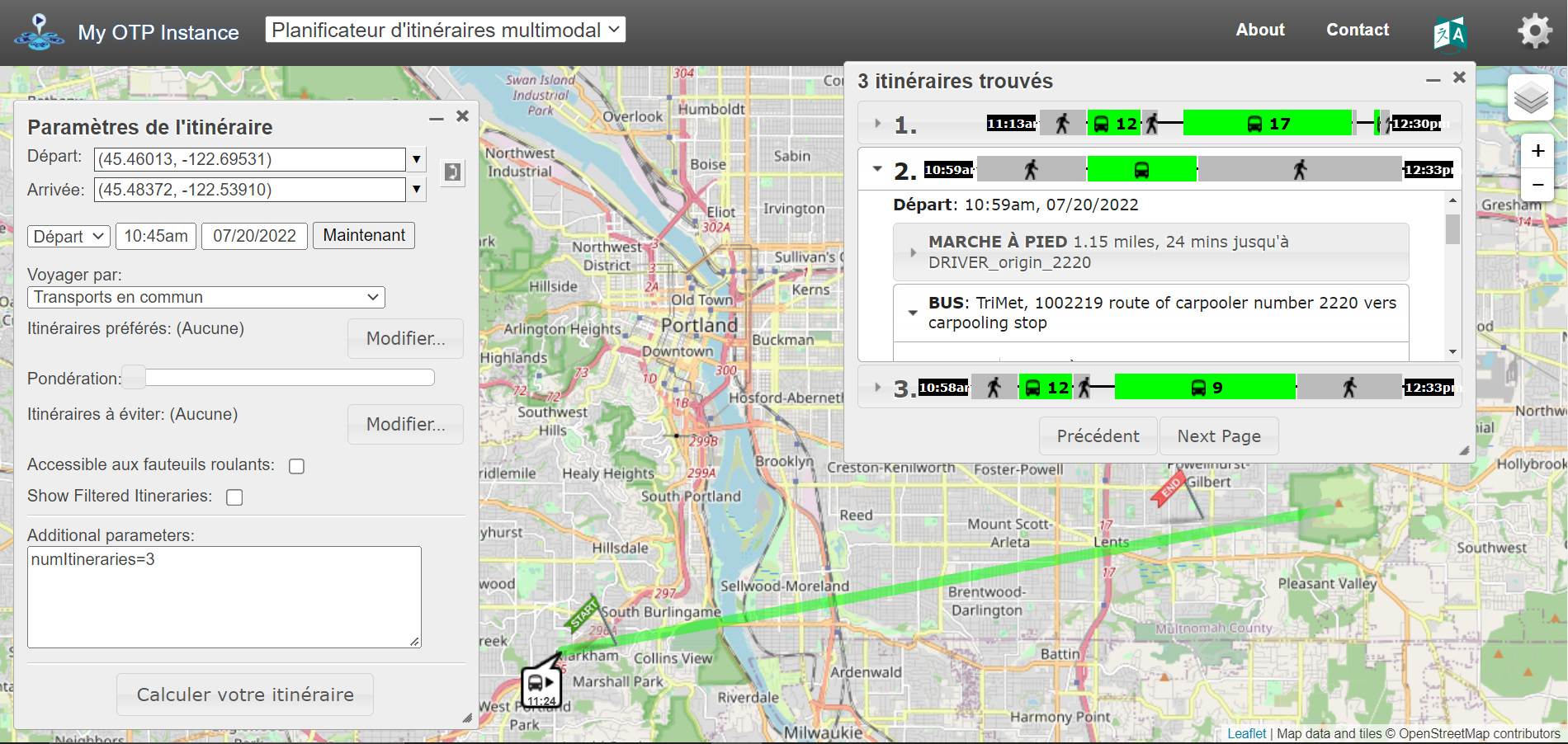}
    \caption{Example of multimodal path as a result of a rider query.}
    \label{fig:otp}
\end{figure}

\subsection{Implementation}
For our numerical results, we wrote a script in Python that generates a scenario $(\mathcal{R},\mathcal{D},\mathcal{PT})$. First, GTFS data are modified adding drivers' journeys $\mathcal{J}^d$ as explained in \S\ref{sec:GTFS} and \S\ref{sec:journeys}. An OTP server is run, taking as input the modified GTFS. The script than generates for each rider $r$ an HTTP request via the API of OTP, as illustrated in \S\ref{sec:rider-journey}. Obviously, for scalability reason, no graphical interface is used at this stage. The paths proposed by OTP are stored in Python variables for further analysis. Although OTP can return multiple alternative paths, we consider only the shortest one for each user. If a driver vehicle reaches its seat capacity, which is a quite rare event,  both this driver and corresponding riders will be discarded from the results. Although a better method to directly discard the routes directly from the OTP request is a work in progress.  

\section{Numerical results}
\label{sec:numerical}

\begin{figure}
    \centering
    \includegraphics[width=\linewidth]{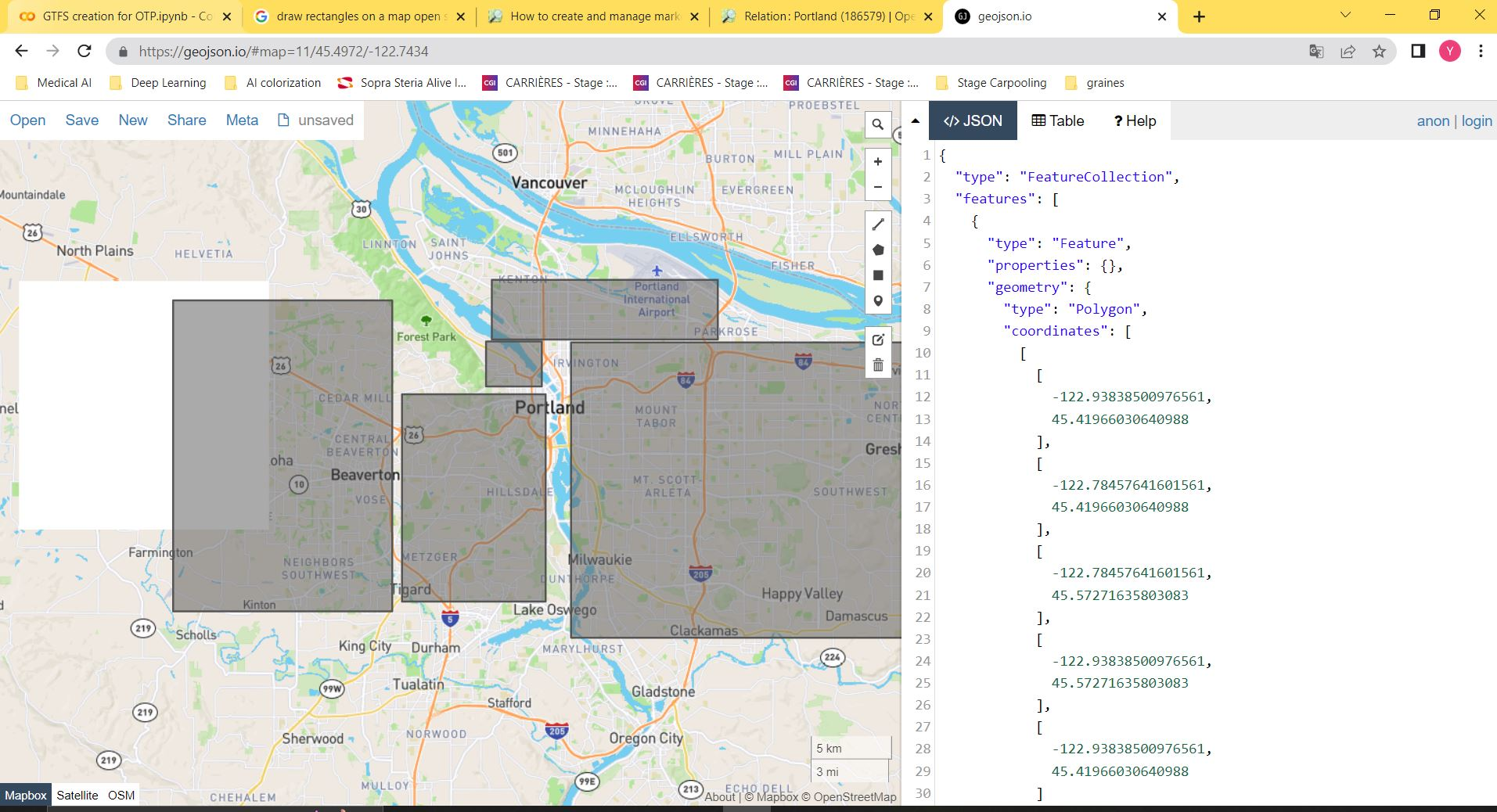}
    \caption{Rectangles used for generating riders and drivers' origins and destinations}
    \label{fig:rectangles}
\end{figure}

We now show the performance of the proposed Integrated System in a large scale metropolitan area. Results confirm the trends we observed in a previous work~\cite{araldo2022pooling}, where however the lack of a proper unifying model for PT and carpooling was limiting the scenarios to have one PT line only.
\todo{aa: Dans la Fig.~\ref{fig:rectangles}, restrainds stp la visualisation à la carte avec les rectangles, en éliminant la barre standard du navigateur ainsi que la partie de script à droite, qui est en tout cas illisible.}

\subsection{Scenario}
For our experiments, we generate drivers and riders origins and destinations by defining several rectangles in the city's map and then choosing one rectangle at random in which we will generate our point in a uniformly random manner.

We generate with this process a density of 4.8 drivers/km²/hour and 8.3 riders/km²/hour. Given that the generation area is 662.47km², we end up with 2848 drivers and 5498 riders. We then impose on drivers a maximum percentage of detour. The simulation takes place from 10:30 AM to 11:30 AM and we build the results statistics from riders that started their journey from 10:45 AM to 11:15 AM. Other numerical values are reported in Table~\ref{tab:notation}. Values that are not specified here are the same as in~\cite[Table~1]{araldo2022pooling}.

For simplicity, we assume all drivers declare their trip at the beginning of the day, before all riders' declaration arrive.

We compare three systems:
\begin{itemize}
    \item \emph{No Carpooling System}, where carpooling is not available, as it is still common in several urban conurbations.
    \item \emph{Current System}, i.e., the status quo, where carpooling exists but it is not integrated with PT: either a user journey is entirely done by carpooling or entirely done within PT.\footnote{In theory, also in this system, a user may autonomously request trips from a certain location to a certain PT stop, then choose to which other stop to go and, from them, reserve another carpooling trip. We ignore such possibility since (i)~in practice it is very unlikely that a user does this quite difficult computation and (ii)~we are interested in the system aspects, instead of user behavior.} 
    \item \emph{Integrated System}, where PT and carpooling are part of the same network: we have conventional lines (buses, subways, tramways, etc.) operating on regular routes and schedules and PoolLines, representing the trips of carpooling drivers. In such system, a user trip request may be satisfied by a multimodal route, combining fixed PT segments and PoolLines.
\end{itemize}

\subsection{Benefits of the Integrated System}
\begin{figure}[t]
    \centering
    \includegraphics[width=1.1\linewidth]{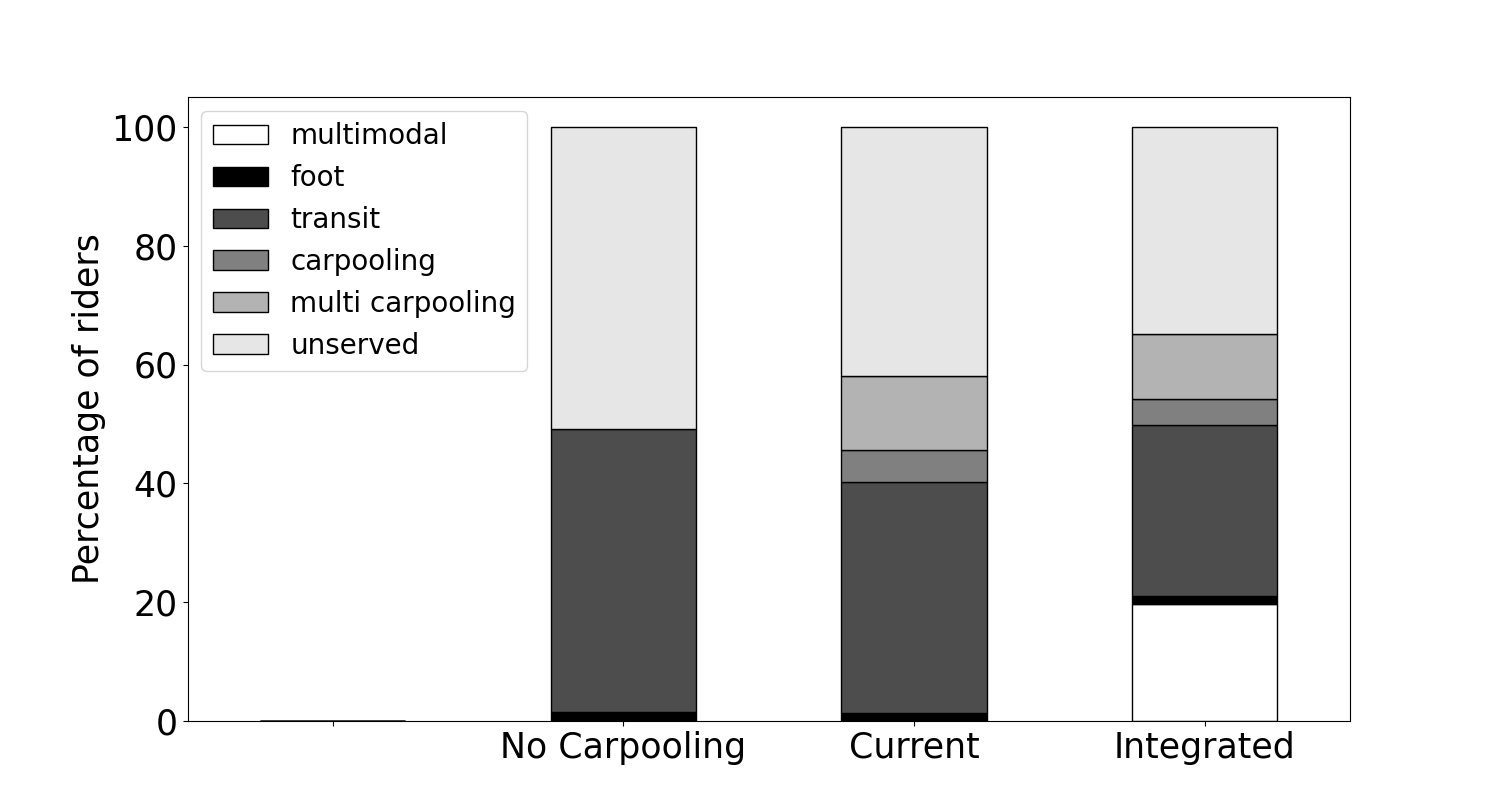}
    \caption{Riders' modal split.}
    \label{fig:modes}
\end{figure}

Fig.~\ref{fig:modes} shows the modal split of the three systems. 

\emph{Unserved} riders' trips are the ones for which no feasible route was found. A route is feasible if (i)~the waiting time (or the sum of all waiting times in case of multiple segment) is less than 45 minutes, (ii)~the total walking distance is less than 2.5 Km and (iii)~the total journey time does not exceed the time the rider would take by walking from origin to destination.

\todo{aa: Dans Fig.\ref{fig:modes}, les couleurs de Multimodal et de Multicarpooling ne sont pas très distinguables. Peux-tu mettre Multimodal en blanc (\#ffffff)? }
\emph{Foot} indicates trips done by walking from the origin to the destination of the rider.  When a rider trip is done by \emph{Carpooling}, she walks to the meeting point indicated by the system controller, she is picked-up by a driver, she is dropped off at another meeting point and, from there, she walks to her destination. \emph{Multi-Carpooling} indicates the path includes multiple carpooling path, with multiple drivers. \emph{Transit} paths include walk and transit segments. \emph{Multimodal} paths include walk, carpooling and transit segments.

Unserved users would be left without travel alternatives and would be obliged to use their private cars. We observe that Carpooling alone reduces unserved users, but such gains are amplified when integrating carpooling and PT. Our proposed Integrated System thus show potential in reducing car-dependency of areas not sufficiently served by conventional PT. Moreover, by considering the set of riders that are only served in the integrated system, meaning riders that would be unserved in the current system. We can showcase by subtracting the distance they would drive in their private cars by the detours taken by drivers that took them on-board in the integrated system that the integrated system saves \emph{6392} kilometers of vehicle travel and by using \cite{ADEME2022}, we can compute the amount of CO2 saved on average which results in \emph{1240} kg of CO2 saved per hour of simulation.

\begin{figure}[t]
    \centering
    \includegraphics[width=0.48\linewidth]{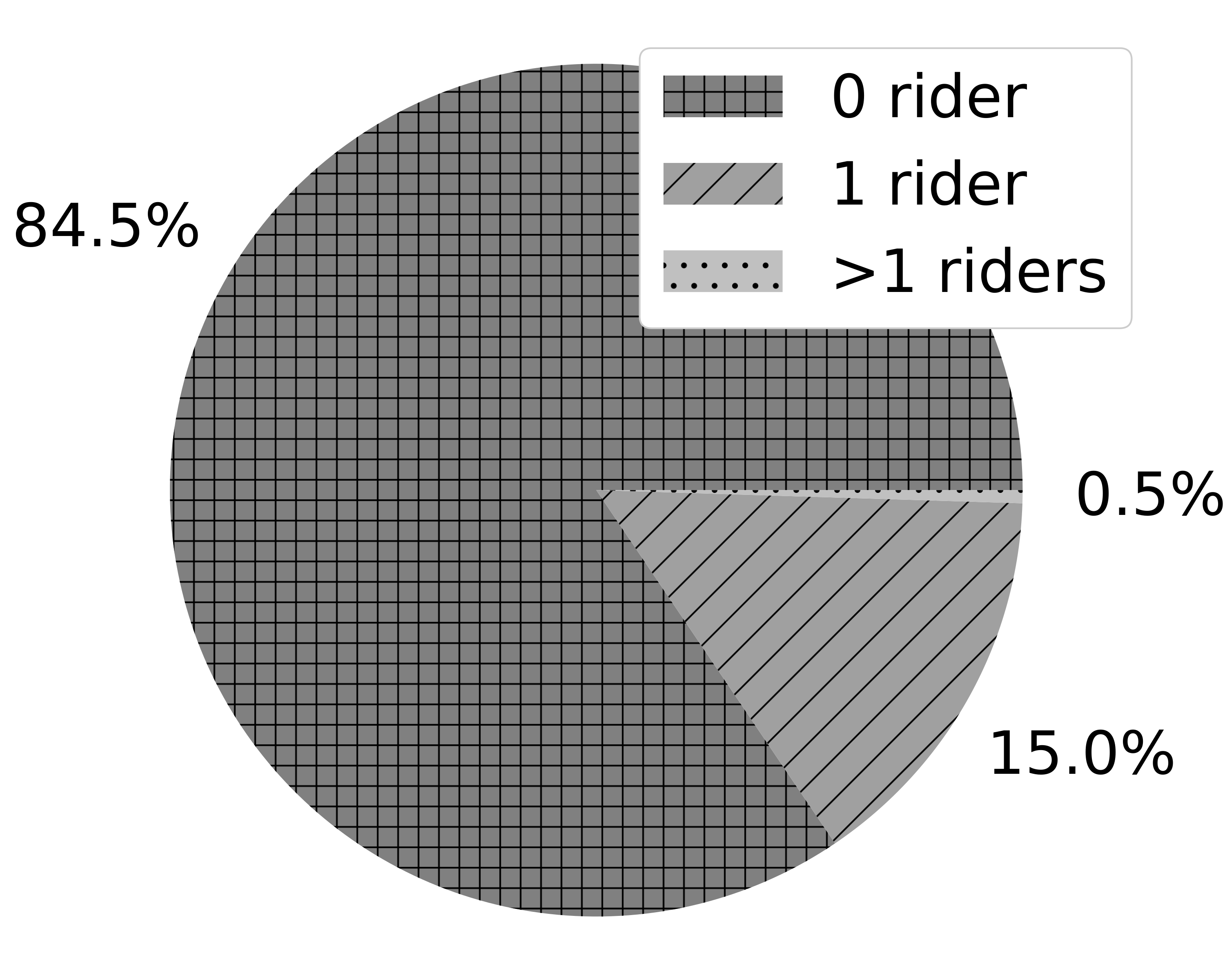}
    \includegraphics[width=0.48\linewidth]{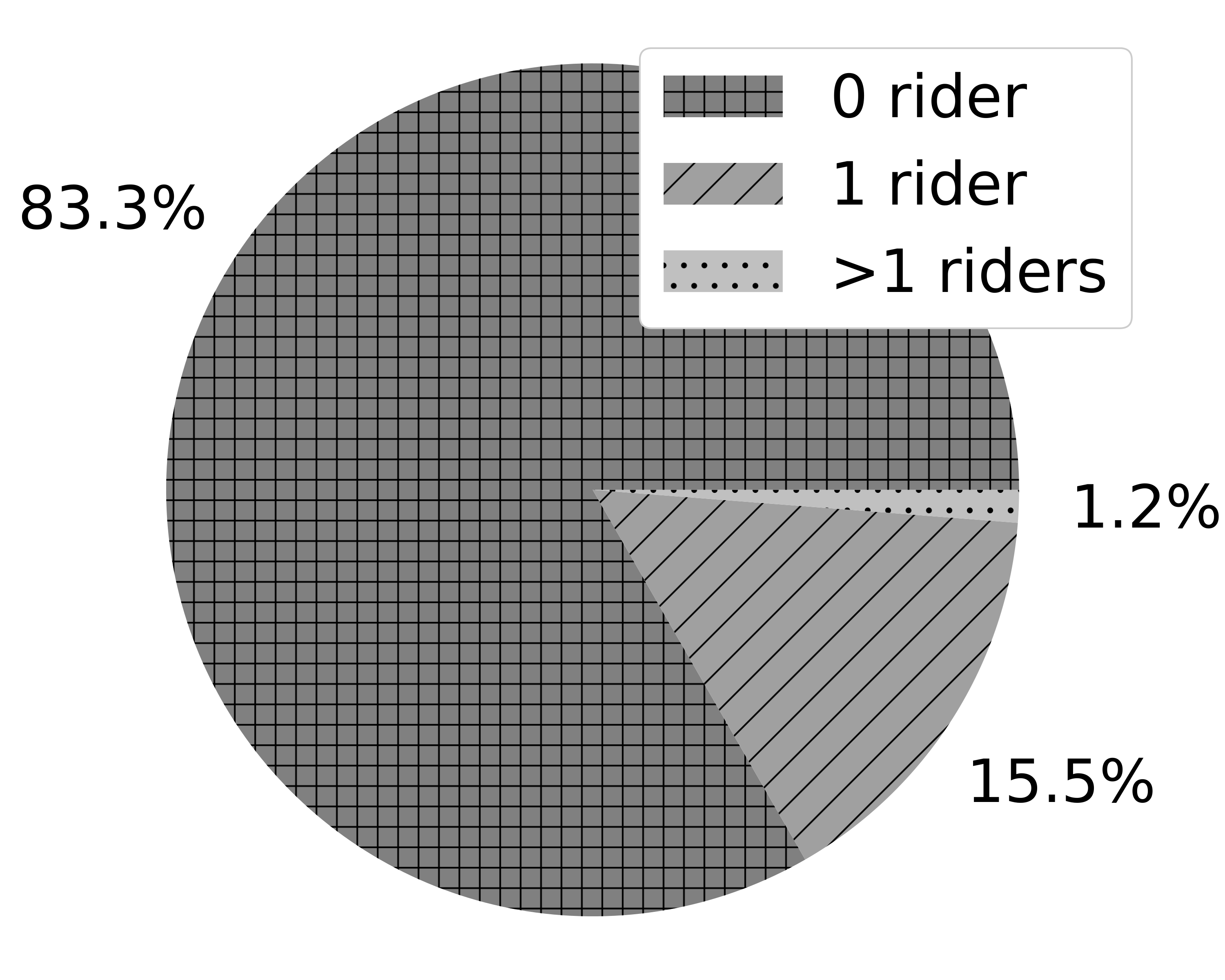}
    \caption{Vehicle occupancy of Current (left) and Integrated (right) systems.}
    \label{fig:occupancy}
\end{figure}

Fig.~\ref{fig:occupancy} shows that our proposed Integrated System enables higher sharing of driver trips. In particular, while trips shared with more than one riders were almost nonexistent in the Current system, they reach 1.2\% in the Integrated system. This can be explained with the low probability of having good matching between origin, destination and departure times of drivers and riders. Having good matching up to (or from to) a PT stop, instead, is much more likely, since stop aggregate demand, thus acting as consolidation points. Higher sharing of driver trips is important from an economic point of view, since it accumulates payments from multiple riders to compensate the same driver trip. 
%
%
%
\begin{figure}[t]
    \centering
    \includegraphics[width=0.6\linewidth]{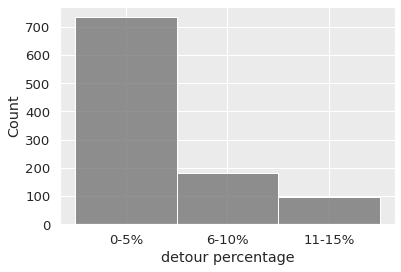}
    \caption{Distribution of drivers' detour (\% in the x-axis).}
    \label{fig:detour}
\end{figure}
\todo{aa: Dans Dif.~\ref{fig:detour}, mets les barres agregées suivantes: [0Km,0Km], [1Km,5Km], [6Km,10Km], [11Km, 15Km]. Comme cela tu peux augmenter bcp la taille des caractères, tout en laissant la taille telle quelle. }
All the above gains are obtained by demanding small detours to drivers as shown in Fig.~\ref{fig:detour}.
\newcommand{\shrinking}{\vspace{-0.2cm}}

\shrinking{}

\section{Conclusion and future work}
\label{sec:conclusion}
We have presented an Integrated System, where conventional fixed PT lines are seamlessly combined with carpooling trips. The underlying approach consists in modeling the entire system in a unified model, where everything is a line: subways, buses, tramways lines are operated following a pre-defined routes; trips of carpooling drivers are also represented by lines (which we call \emph{PoolLines}). Such unified model is stored in a GTFS database, which is then used by off the shelf routing software (OpenTripPlanner in our case) to compute multimodal routes. The benefits of such an Integrated System with respect to the Current system, where carpooling and PT are operated separately, are lower levels of unserved riders and higher sharing of drivers' trips.

Some heavy assumption make our work preliminary, as the fact that rider demand is inelastic, the omission of explicit model of private car as available alternative to riders, the lack of congestion models and the lack of payment models.

\shrinking{}

\section*{Aknowledgement}
Supported by The French ANR research
project MuTAS (ANR-21-CE22-0025-01).

\shrinking{}
\bibliographystyle{abbrv}
\bibliography{references}


\end{document}